\documentclass[12pt]{article}

\usepackage{amsmath,amssymb}
\usepackage[nosort]{cite}
\def\sl2R{sl(2,\mathbb{R})}
\def\SL2R{SL(2,\mathbb{R})}

\def\tr{{\rm Tr}}
\def\half{\frac{1}{2}}

\def\thefootnote{\fnsymbol{footnote}}
\def\11{\mbox{$1$}}

\renewcommand{\thefootnote}{\alph{footnote}}

\newcommand{\rref}[1]{(\ref{#1})}
\newcommand{\beqn}{\begin{equation}}
\newcommand{\eeqn}{\end{equation}}
\newcommand{\beqarr}{\begin{eqnarray}}
\newcommand{\eeqarr}{\end{eqnarray}}

\newcommand{\matc}{\begin{array}{c}}
\newcommand{\matcc}{\begin{array}{cc}}
\newcommand{\matccc}{\begin{array}{ccc}}
\newcommand{\matcccc}{\begin{array}{cccc}}
\newcommand{\emat}{\end{array}}

\newcommand{\IH}{\relax{\rm I\kern-.18em H}}
\newcommand{\IR}{\relax{\rm I\kern-.18em R}}
\newcommand{\IK}{\relax{\rm I\kern-.18em K}}
\newcommand{\II}{\hbox{\rm 1\kern-.28em I}}
\newcommand{\Is}{\relax{\rm 1\kern-.35em 1}}
\begin{document}

\begin{titlepage}

October 2005 

~ \hfill  RU-05-4-B

~ \hfill CCNY-HEP-05/6

\begin{center}
\vskip .2in
\renewcommand{\thefootnote}{\fnsymbol{footnote}}
{\LARGE Fractional quantum Hall effect on the two-sphere:\\
\vskip .1in
a matrix model proposal}
\vskip .45in
Bogdan Morariu$^{a}$ and
Alexios P. Polychronakos$^{b}$
\vskip .45in

{\em $^{a}$Physics Department, Rockefeller University \\
        New York, NY 10021, USA \\
        {\rm {\small morariu@summit.rockefeller.edu}}
\vskip .1in
        $^{b}$Physics Department, City College of the CUNY \\
        New York, NY 10031, USA \\
        {\rm {\small alexios@sci.ccny.cuny.edu}}
}
\end{center}
\vskip 0.45in
\begin{abstract}
We present a Chern-Simons matrix model describing the fractional quantum Hall
effect on the two-sphere. We demonstrate the equivalence of our proposal 
to particular restrictions of the Calogero-Sutherland model, reproduce the 
quantum states and filling fraction and show the compatibility of our result
with the Haldane spherical wavefunctions.
\vskip 0.45in
~ \hfill PACS numbers: 11.10.Nx; 73.43.-f

\end{abstract}

\end{titlepage}

\newpage
\renewcommand{\thepage}{\arabic{page}}

\setcounter{page}{1}
\setcounter{footnote}{0}


\section{Introduction}
\label{Intro}
In paper \cite{Susskind:2001fb}, Susskind proposed to describe the
fractional quantum Hall fluid using noncommutative Chern-Simons
field theory. (For some recent reviews of noncommutative field theory
see~\cite{Seiberg:1999vs, Harvey:2001yn, Douglas:2001ba, Szabo:2001kg}.)
In \cite{Susskind:2001fb}, the inverse filling fraction is identified with the
Chern-Simons level, and the coordinates of electrons are elevated to
matrices which are essentially the spatial components of the
noncommutative gauge field. Just as in the ordinary Chern-Simons, the
spatial components of the gauge field have nontrivial commutation
relations and as a consequence the electron coordinates are
automatically projected to the lowest Landau level.  The
model~\cite{Susskind:2001fb} describes an infinite number of electrons
and, modulo gauge invariance, has only one state.

Since the original proposal~\cite{Susskind:2001fb}, significant work
has been done extending the model and relating it to previous
descriptions of the fractional quantum Hall effect. A matrix model
describing finite size Hall droplets and supporting boundary
excitations was introduced in~\cite{Polychronakos:2001mi}. The model
naturally contains quasiholes and one can derive their quantized
charge and the quantization of the inverse filling fraction. The relation
between the states of the model~\cite{Polychronakos:2001uw} and the
Laughlin states was further explored in~\cite{Hellerman:2001rj,
  Karabali:2001xq, Hansson:2001kt, Cappelli:2004xk}. 
See also~\cite{Polychronakos:2001uw} for
an extension of the model describing the Hall fluid on a cylinder
and~\cite{Morariu:2001qa} for an attempt to describe a multi-layered
Hall fluid.

The construction of the corresponding matrix models on compact spaces,
however, presents special challenges. In particular, no Chern-Simons
like action exists for the two-sphere. Alternative approaches have
been suggested in \cite{Chen:2003ic} where two complex matrices were
used to parametrize the coordinates of the electrons and in
\cite{Berenstein:2004hw} where a fermionic matrix model was introduced
to describe a Hall fluid at filling fraction one. See
also~\cite{Ghodsi:2005ks} for further discussions. The lack of a
Chern-Simons type action is surprising, since Laughlin's treatment
\cite{Laughlin:1983fy} of the fractional quantum Hall effect was
easily extended by Haldane \cite{Haldane:1983xm} to the two-sphere.
The matrix model describing this situation is expected to reproduce
features of the ``fuzzy sphere'', at least for the fully filled case.
(For early work on fuzzy spheres and some further discussions see
\cite{Madore:1991bw, Carow-Watamura:1996wg, Klimcik:1997mg,
  Gratus:1997rq, Morariu:1997vu, Nair:2000ii, Morariu:2004aw}.)
  
In this paper we shall try to
rectify this situation by proposing such a matrix model
using stereographic coordinates on the sphere and a boundary field
similar to the one in the finite planar model of \cite{Polychronakos:2001mi}.
Furthermore,
we show, using two different natural changes of variables, that our
model is equivalent to either a Calogero \cite{Calogero:1970nt} or a
Sutherland \cite{Sutherland:1971kq} integrable
model. Some restrictions exist on the values of the integrals of
motion but these are consistent with the dynamics, even when we
perturb it by reasonable hamiltonians.  We also show that
once these restrictions are imposed, the Calogero or Sutherland models
acquire a new $SO(3)$ symmetry corresponding to three dimensional
rotations of the spherical Hall fluid. 

The plan of the paper is as follows. In Section \ref{Lfluid} we
present the action for a charged (lagrangian) fluid on a two-sphere in
a strong magnetic field. In Section \ref{MModel} we introduce a
matrix model whose classical limit is the fluid model. We discuss
ordering issues and show that the model admits the fuzzy sphere as a
solution. In Section \ref{FMModel} we present our second, finite model
which incorporates a boundary field. In Section \ref{Equivalence} we
show using the equivalence of the model to either a Calogero or a 
Sutherland model with
some restrictions on the values of the integrals of motions.
In Section \ref{Quantization} we perform the quantization of the model
and compare its states with the corresponding
Haldane states on the sphere. Finally, we conclude with
a set of comments and open questions in Section \ref{remarks}.

\section{Lagrangian fluid model}
\label{Lfluid}

As discussed in~\cite{Bahcall:1991an}, and more recently reviewed
in~\cite{Susskind:2001fb}, one can describe the Hall fluid using a
lagrangian fluid model. The electrons of the fluid are described in
the Lagrange formulation by their spatial coordinates $x_a (y_1,y_2)$
which are functions of the particle-fixed coordinates $y_a$ ($a=1,2$
for the Hall system).  Assuming that the body-fixed coordinates
represent a reference state of constant particle density $\rho_0$ and
charge density $e \rho_0$, the lagrangian of the model is given by
\begin{equation}
  {\cal L} =
eB\rho_0\int
d^2y
\left[\,\frac{1}{2}\,\epsilon^{ab}x_a \dot{x}_b + 
\frac{A_0}{2 \pi \rho_0} \left(\{x_1,x_2\}-1\right)
\right]~,
\end{equation}
where $\{x_1,x_2\}$ are the Poisson brackets on the manifold $y_1,y_2$:
\begin{equation}
\{x_1,x_2\} = \epsilon^{ab}\partial_a x_1 \partial_b x_2
= \frac{\partial x_1}{\partial y_1}\frac{\partial x_2}{\partial y_2}
- \frac{\partial x_1}{\partial y_2}\frac{\partial x_2}{\partial y_1} ~.
\end{equation}
The
first term in $\cal L$ is the coupling of the charged fluid to a constant
magnetic field $B$. A constant density for the fluid equal to $\rho_0$
is enforced by the
lagrange multiplier $A_0$. For more details and the motivation for this
model see the discussion in~\cite{Susskind:2001fb}.

We use the same approach to treat the Hall fluid on a two-sphere.
The gauge field for a uniform magnetic field of
strength $B$ on a two-sphere can be written in stereographic
coordinates (up to a gauge transformation) as
\begin{equation}
 i2BR^2(1+z\bar{z})^{-1}\bar{z}dz~.
\label{monopole}
\end{equation}
(From now on we put $e=1$ for simplicity.)
Near the north pole of the sphere the stereographic coordinate $z$
becomes $\sim (x+iy)/2R$ in terms of local Cartesian coordinates,
while $z=\infty$ represents the south pole and the gauge potential
has a Dirac-string singularity there. 

The lagrangian description of the Hall fluid is given by a map
$z(\sigma,\bar{\sigma})$ where both $z$ and $\sigma$ are stereographic
coordinates on the two-sphere.
We can still pick the lagrangian parametrization
$(\sigma,\bar{\sigma})$ such that it corresponds to a reference
configuration of uniform particle density $\rho_0$ in $\sigma$ space
\begin{equation}
  dN=\rho_0\frac{2R^2}{(1+\sigma\bar{\sigma})^2}d\sigma
  d\bar{\sigma}~.
\label{sigmaDens}
\end{equation}

The coupling of the charged fluid to the magnetic field~\rref{monopole}
is given by
\begin{equation}
   {\cal L'} =  i2BR^2
\int \rho_0 \frac{2R^2 d\sigma d\bar{\sigma}}{(1+\sigma\bar{\sigma})^2}
(1+z\bar{z})^{-1}\dot{z}\bar{z} ~.
\end{equation}
As in \cite{Susskind:2001fb} we consider a limit where the coupling to the
magnetic field dominates the standard quadratic kinetic and potential
terms, and the latter terms will be ignored in what follows (a classical
lowest Landau level reduction \cite{Dunne:1989hv}).

As in the planar case, we want to restrict the model to configurations
of uniform density. Then the infinitesimal particle number in $z$
space must be given by 
\begin{equation}
  dN=\rho_0 \frac{2R^2}{(1+z\bar{z})^2}dz d\bar{z}~,
\label{zDens}
\end{equation}
with the same $\rho_0$ as in~\rref{sigmaDens}.  Thus
the map $z(\sigma,\bar{\sigma})$ must be area preserving, and we have
\begin{equation}
  \label{AreaPreserv}
  (1+z\bar{z})^{-2} \left(
\frac{\partial z}{\partial \sigma}
\frac{\partial \bar z}{\partial \bar\sigma} -
\frac{\partial z}{\partial\bar \sigma}
\frac{\partial \bar z}{\partial \sigma}
\right)=
 (1+\sigma\bar{\sigma})^{-2}~.
\end{equation}
We can rewrite this using the Poisson bracket on the sphere
\begin{equation}
  i\{F,G\}=(1+\sigma\bar{\sigma})^2 
\left(
\frac{\partial F}{\partial \sigma}
\frac{\partial G}{\partial \bar\sigma} -
\frac{\partial F}{\partial\bar \sigma}
\frac{\partial G}{\partial \sigma}
\right)~.\label{symp}
\end{equation}
Note that the Poisson tensor is just the inverse of the symplectic
(area) $2$-form in stereographic coordinates. Using \rref{symp} we
write the constraint~\rref{AreaPreserv} as
\begin{equation}
  (1+z\bar{z})^{-2} \{z,\bar{z}\} +i =0~.\label{con}
\end{equation}
We can obtain~\rref{con} directly from a lagrangian by introducing a
multiplier $A_0$
\begin{equation}
  \label{fluid}
  {\cal L} =  i2BR^2 
\int \rho_0
\frac{2 R^2 d\sigma d\bar{\sigma}}{(1+\sigma\bar{\sigma})^2}
\left\{
(1+z\bar{z})^{-1}\dot{z}\bar{z} +
\frac{A_0}{2\pi\rho_0}\left[(1+z\bar{z})^{-2} \{z,\bar{z}\} +i
\right]
\right\}~.
\end{equation}
As the notation suggests, $A_0$ can be interpreted as a gauge
field. Indeed, defining the covariant derivative
\begin{equation}
  D_0 z=\dot{z}+\theta\{A_0,z\}~,
\end{equation}
where $\theta=(2\pi\rho_0)^{-1}$, we can combine the first two terms in
the lagrangian and (up to a total derivative) we have
\begin{equation}
  \label{fluiD}
  {\cal L} =  i2eBR^2
\int \rho_0
\frac{2 R^2 d\sigma d\bar{\sigma}}{(1+\sigma\bar{\sigma})^2}
\left\{
(1+z\bar{z})^{-1}D_0 z\bar{z} +
\frac{i A_0}{2\pi\rho_0} 
\right\}~.
\end{equation}
The gauge group corresponding to $A_0$ is the infinite dimensional group
of area preserving diffeomorphisms.

\section{Noncommutative model}
\label{MModel}

The noncommutative version (matrix model) of the fluid model on the
sphere is obtained by promoting the field $z(\sigma,\bar{\sigma})$
into a matrix $z$ and turning the two-dimensional integration into a
matrix trace
\begin{equation}
  \int \frac{2R^2 d\sigma d\bar{\sigma}}{(1+\sigma\bar{\sigma})^2}
\rightarrow
2\pi\theta \tr~.
\end{equation}
The dimension of the matrix $N$ will
represent the number of particles.

In writing the appropriate action, we face issues of matrix ordering,
since the matrices $z$, $z^\dagger$ and $\dot z$ do not commute.
We shall use a lagrangian $L$ with an ordering in which $z$ and
$z^\dagger$ alternate; specifically the kinetic term is given by:
\begin{equation}
{\cal L'} (z,z^{\dagger})=
i 2B R^2 \,\tr\left[(1+zz^{\dagger})^{-1}
\dot{z}z^{\dagger}
\right]~.
\end{equation}
The same ordering also appears in the kinetic terms of certain
nonlinear sigma models discussed in \cite{Binetruy:1984yx,
  Gaillard:1997zr}.
Such an ordering ensures that the above lagrangian properly describes
a model on the sphere. As in the single-particle (scalar) case,
this requires that the canonical structure obtained by the above
lagrangian be invariant under a set of $SO(3)$ rotations. Such
transformations are a subgroup of the full group $SL(2,\mathbb{C})$ of modular
transformations defined, as in the scalar case, by
\begin{equation}
z' =
(Az+B)(Cz+D)^{-1}~, \label{modular}
\end{equation}
with $A,B,C,D$ complex numbers satisfying $AD-CB=1$.
Specifying $SO(3)$\,, or equivalently $SU(2)/\mathbb{Z}_2$
transformations, amounts to
choosing a unitary matrix of the above coefficients:
\begin{equation}
\left(
\begin{array}{cc}
A & B\\
C & D
\end{array}
\right)
\left(
\begin{array}{cc}
A & B\\
C & D
\end{array}
\right)^\dagger = \mathbb{I}~. \label{so3}
\end{equation}
Note also that an overall phase in this matrix is irrelevant in the
transformations~\rref{modular}\,.
Then we can show that
the above lagrangian changes by a total derivative,
\begin{equation}
{\cal L'}(z',{z'}^{\dagger})={\cal L'}(z,z^{\dagger}) +
\frac{d}{dt}\left\{
i 2B R^2 \,
\tr\left[{\rm ln}(Cz+D)\right]\right\}
\end{equation}
and therefore the canonical form remains invariant. Note that the
chosen ordering is crucial for this property\footnote{Stereographic
  coordinates have been used previously to study fuzzy spheres
  in~\cite{Alexanian:2000uz, Iso:2001mg}. However, these papers use a
  slightly different definition of $z$ in terms of
  ``Cartesian coordinates'' than is used here. As a
  consequence $z$ does not transform by fractional transformations.}.

In infinitesimal form, the relations~\rref{so3} imply
\begin{equation}
\left(
\begin{array}{cc}
A & B\\
C & D
\label{infinitesimal}
\end{array}
\right)
 \approx \mathbb{I}+\frac{i}{2}\,
\left(
\begin{array}{cc}
a & b\\
\bar{b} & -a
\end{array}
\right)~,
\end{equation}
with $\bar{a} = a$. This corresponds to the
transformation
\begin{equation}
\delta z = i\,(a z + \frac{1}{2}\,b - \frac{1}{2}\,\bar{b} z^2)
\label{trans}
\end{equation}
representing infinitesimal rotations. By Noether's theorem, the
conserved charges corresponding to the above rotations are
\begin{equation}
\left\{
\begin{array}{lll}
J_{3} & = & B R^2\,\tr\left[(1+zz^\dagger)^{-1} + 
(1+z^\dagger z)^{-1} -1\right]~,\\
J_{+} & = & 2 B R^2\,\tr\left[(1+zz^{\dagger})^{-1}z\right]~, \\
J_{-} & = & 2 B R^2\,\tr\left[z^{\dagger}(1+zz^{\dagger})^{-1}\right]~.
\end{array}
\right.
\label{Jclass}
\end{equation}
These are the generators of the transformation (\ref{infinitesimal})
and consequently their Poisson brackets satisfy the $SU(2)$ rotation
algebra
\begin{equation}
\begin{array}{lll}
\left\{ J_{\pm}, J_3\right\} & = &  \pm \,i\, J_{\pm}~,\\
\left\{ J_{-},J_{+}\right\} & = & i 2 \,J_3~.
\end{array}
\label{JJ} 
\end{equation}
This result can be directly verified by using the canonical Poisson
brackets implied by the lagrangian, from which we read off the Poisson
structure
\begin{equation}
\{ z_{a_1 b_1} , [z^\dagger (1+z z^\dagger )^{-1} ]_{a_2 b_2} \} =
\frac{1}{i2BR^2} \delta_{a_1 b_2} \delta_{a_2 b_1}~.
\end{equation}
The above can be written in a more convenient index-independent
notation, denoting by $1$ and $2$ the spaces in which indices ($a_1 ,
b_1$) and ($a_2 , b_2$) act respectively, and by $T_{12}$ the
operator exchanging the two spaces. In this notation we have
\begin{equation}
\{ z_1 , z_2^\dagger (1_2 + z_2 z_2^\dagger )^{-1} \} =
\frac{1}{i2BR^2} T_{12}~.
\end{equation}
The Poisson brackets of $z$ and $z^\dagger$ derive from above as:
\begin{equation}
\{ z_1 , z_2^\dagger \} = \frac{1}{i2BR^2} (1_1 + z_1 z_1^\dagger )
\, T_{12} \, (1_1 + z_1^\dagger z_1 )~.
\label{canon}
\end{equation}
Using (\ref{canon}) we can show that the $SU(2)$ Poisson brackets
(\ref{JJ}) hold. (More details of the derivation are given in
the appendix.)

The ordering of the matrices inside the trace in the definition of
$J_\pm$ and $J_3$ is essentially fixed by the alternating $z$ and 
$z^\dagger$ rule, up to cyclic permutations.
The first two terms in the definition of $J_3$ are actually
equal inside the trace. The particular ordering was used for the
following reason: the matrices
\begin{equation}
\left\{
\begin{array}{lll}
X_3 & = & R [ (1+zz^\dagger )^{-1} + (1+z^\dagger z)^{-1} -1 ]~,\\
X_+ & = & 2R (1+zz^{\dagger})^{-1} z ~,\\
X_- & = & 2R z^{\dagger}(1+zz^{\dagger} )^{-1}~,
\end{array}
\right.
\label{XXX}
\end{equation}
transform to each other as a vector under the transformation (\ref{trans}),
as can be directly checked. They are, therefore, prime candidates for the
(Cartesian) matrix coordinates on a sphere. This point will be relevant
in what follows.

The matrix model presented above describes a ``fuzzy fluid'' consisting
of $N$ particles on the sphere in the presence of a constant
magnetic field. We would also like to incorporate the incompressibility
condition that the density on the sphere of this fluid remains
constant. Just as in the planar case, this can be implemented by
gauging the time derivative
\begin{equation}
\dot{z} \to \dot{z} - i [ A_0 , z ]~,
\end{equation}
with $A_o$ a hermitian matrix gauge field, and adding the term 
$B \theta \tr A_0$ in the lagrangian. The Gauss law obtained from
varying $A_0$ reads
\begin{equation}
[ z , z^\dagger (1+z z^\dagger )^{-1} ] \equiv
(1+z^\dagger z)^{-1} - (1+z z^\dagger)^{-1}
= \frac{\theta}{2R^2}~.
\label{nacom}
\end{equation}
This is the noncommutative (matrix) analog of the fluid condition
that fixes the density on the sphere to 
\begin{equation}
\rho_0 = \frac{1}{2\pi \theta}
\end{equation}
(given that $z \sim x/2R$).

The above commutation relations~\rref{nacom} are 
very suggestive. Consider the three
\mbox{``coordinate''} matrices $X_\pm$ and $X_3$ as defined in (\ref{XXX}).
We can show that the commutation relations~\rref{nacom} are 
formally equivalent
to $SU(2)$ commutation relations among the coordinates up to a factor
of $\theta/R$:
\begin{equation}
[X_3 , X_\pm ] = \pm \frac{\theta}{R} X_\pm ~,~~~
[X_+ , X_- ] = \frac{\theta}{R} 2 X_3~.\label{fuzzy}
\end{equation}
Moreover, the sum of squares of the above three matrices is
\begin{eqnarray}
X_+ X_- + X_- X_+ + 2 X_3^2 &=& 2 R^2 \left\{ 1 - [ (1+z^\dagger z)^{-1}
- (1+z z^\dagger )^{-1} ]^2 \right\} \nonumber \\
&=& 2R^2 \left[ 1 - \left(\frac{\theta}
{2 R^2} \right)^2 \right]~,
\label{Xsquare}
\end{eqnarray}
the last equality being valid upon use of the Gauss law. We see that
the above matrices become coordinates of a fuzzy sphere of radius $R$
and noncommutativity parameter $\theta$. Identifying (\ref{Xsquare})
with ($\theta^2/R^2$ times) the quadratic Casimir $j(j+1)$
also implies a quantization of $\theta$ according to
\begin{equation}
\theta = \frac{R^2}{j+\half} = \frac{2 R^2}{N}
\label{thecon}
\end{equation}
for $N=2j+1$, as is standard in noncommutative spheres. So this would 
correspond to a completely filled sphere, in analogy to the completely
filled plane of the standard (planar) noncommutative Chern-Simons
model.

Unfortunately, the matrix relation (\ref{nacom}) is inconsistent.
By tracing both sides of the Gauss law we see that, just as in the
planar case, no finite-dimensional
matrices can satisfy it. Unlike the planar case, however, it admits
no infinite-dimensional representations either. To see this, rewrite
the above commutator as
\begin{equation}
[ z , z^\dagger z (1+ z^\dagger z)^{-1} ] =
 \frac{\theta}{2R^2} z~.
\end{equation}
So $z$ and $z^\dagger$ act as lowering and raising operators,
respectively, for both $z z^\dagger (1+z z^\dagger )^{-1}$ and
$z^\dagger z (1+z^\dagger z)^{-1}$. (Assuming $\theta>0$; else the
role of $z$ and $z^\dagger$ is reversed.) But the spectrum of these
last two operators is bounded from above by $1$; therefore, there
must be a highest state, annihilated by $z^\dagger$. On this state,
however, the relation (\ref{nacom}) implies that 
$z^\dagger z (1+z^\dagger z)^{-1}$ has a negative eigenvalue,
which is also inconsistent.

Nevertheless, it is remarkable that the fuzzy sphere 
algebra~\rref{fuzzy} can be formally extracted from the commutation
relations~\rref{nacom}. Note further that we should not expect the
matrix elements of $z$ and $z^\dagger$ to be finite. In the
large $N$ limit such finite matrices would map to a nonsingular 
function $z(\sigma,\bar{\sigma})$ from $S^2$ to $\mathbb{C}$. 
However, for the fully filling Hall fluid $z(\sigma,\bar{\sigma})$
is always singular at the south pole. 
Strictly speaking $z$ should be
considered as a local coordinate map of a functions from $S^2$ to
$S^2$\, of winding number one. 
Just as for the classical case it is useful
to work with singular functions, in the matrix case
we should work
with the algebra generated by $z$ and $z^\dagger$ modulo the
relations~\rref{nacom}. While this algebra, as shown above, has no
unitary representations, it contains a subalgebra (generated by
$J_{\pm}, J_2$) that has finite dimensional unitary representations.

This interpretation is useful if we are willing to work with
only the equation of motion. For the fully filled sphere this is
perhaps enough, since modulo gauge transformations there exists only
one state. For partially filled states, however, and for the purposes
of quantization, this description fails.
We conclude that the na\"ive gauged matrix model on the sphere cannot
stand, unlike the planar case.
This is due to the compactness of space: a Laughlin
(or rather Haldane) state on the sphere has a finite
number of particles and would require a finite-dimensional matrix,
which is inconsistent with the commutation relation (\ref{nacom}).

\section{Finite matrix model}
\label{FMModel}

In what follows we introduce a modified model containing a boundary
field, just as in the planar or cylindrical case. As a result the
model can describe Hall fluid states that are only partially
filling the two-sphere with the matrices
$z$ and $z^\dagger$ having finite dimensional representations.

The full spherical boundary matrix model lagrangian reads
\begin{eqnarray}
{\cal L}(z,z^{\dagger})&=&
i B \,\tr\left[2R^2 (1+zz^{\dagger})^{-1}
\dot{z}z^{\dagger}
- i A_o (2R^2 [z,z^{\dagger}(1+zz^{\dagger})^{-1}]- \theta) \right]
\nonumber\\
&+& \Psi^\dagger (i \dot{\Psi} + A_o \Psi )~,
\end{eqnarray}
with $\Psi$ a complex column $N$-vector. In the limit $z \ll 1$ this
reduces to the planar finite quantum Hall matrix model, upon identifying
$X + i Y = 2R z$. This model still admits the transformation
$z' = (Az+B)(Cz+D)^{-1}$ as a symmetry, with $A_0$ and $\Psi$ transforming
trivially, and thus properly describes a spherical system. A hamiltonian
term $-\tr V(z,z^\dagger )$ can also be added, just as in the planar
case, representing a potential that would tend to concentrate the
droplet and assign different energies to different states.

The classical analysis of the dynamics of the model parallels the
one of the planar case. The Gauss law constraint, now, reads
\begin{equation}
2BR^2 [ z , z^\dagger (1+z z^\dagger )^{-1} ] + \Psi \Psi^\dagger
= B \theta~.
\label{newcom}
\end{equation}
This equation admits finite-dimensional representations, under some
conditions for $R$, $B$ and $\theta$. We shall demonstrate the solution
corresponding to a circular droplet centered around the north pole,
a sort of ground state configuration.

Tracing (\ref{newcom}) we obtain
\begin{equation}
\Psi^\dagger \Psi = BN \theta~.
\end{equation}
By a gauge rotation we can bring the $N$-vector $\Psi$ to the form
\begin{equation}
\Psi_n = \delta_{n,N} \sqrt{BN\theta} ~,~~~ n = 1, \dots N~.
\end{equation}
In this basis we shall choose $z$ and $z^\dagger$ to act as lowering
and raising matrices
\begin{equation}
z_{mn} = a_m \delta_{m+1,n} ~,~~~ m,n = 1, \dots N
\end{equation}
determined by the $N-1$ constants $a_1, \dots a_{N-1}$ ($a_N$
does not appear since $\delta_{N+1,n} = 0$). We have
\begin{equation}
(z^\dagger z)_{mn} = |a_{m-1} |^2 \delta_{mn} ~,~~~
(z z^\dagger )_{mn} = |a_m |^2 \delta_{mn}~,
\end{equation}
with the convention $a_0 = a_N = 0$. Putting 
$b_n = |a_n |^2/(1+|a_n |^2 )$ the
commutation relation (\ref{newcom}) gives
\begin{equation}
b_n - b_{n-1} = \frac{\theta}{2R^2} (1-N\delta_{n,N} )~,
\end{equation}
which admits as solution
\begin{equation}
b_n = n \frac{\theta}{2R^2} ~, ~ n<N, ~~~ b_N = 0~.
\end{equation}
The ``problematic'' top state $n=N$, which was giving the inconsistency
in the absence of the boundary field, is now consistent. Finally, the
$a_n$ can be determined from the $b_n$ as
\begin{equation}
| a_n |^2 = \frac{\theta n}{2R^2 - \theta n} ~,~ n<N ~,~~~ a_N = 0~.
\end{equation}
The phases of $a_n$ are gauge degrees of freedom and can be chosen zero.

Consistency requires that the right-hand side above be non-negative. This
imposes the constraint
\begin{equation}
\theta \le \frac{2R^2}{N-1}~.
\label{thenewcon}
\end{equation}
To see what this means, we rewrite it as
\begin{equation}
2\pi \theta (N-1) \le 4\pi R^2~.
\end{equation}
The right hand side is the area of the sphere. $2\pi \theta$ is the
``area quantum'' occupied by each particle, or rather, the area excluded
from occupation by the remaining particles. Thus, positioning $N$
particles on the sphere requires an area of at least (N-1) such
quanta.

The limiting case when $\theta (N-1) = 2 R^2$ corresponds to
positioning the first particle on the north pole, and positioning
subsequent ones on ``fuzzy rings'' at a vertical intervals of
$2R/(N-1)$, with the last particle barely squeezing at the
south pole.  This represents a fully filled sphere. Such a state is,
actually, rotationally invariant, as can be seen from the fact that
the generators $J_\pm , J_3$ vanish in this case, or from the fact
that the $SO(3)$ transformations are equivalent to $SU(N)$
conjugations (gauge transformations) of $z$ in this case. This is the
classical noncommutative analog of the fully filled Laughlin-Haldane
state on the sphere. (Note that one matrix element of $z$ becomes
singular in this case.)

The constraint (\ref{thenewcon}) above differs from the one in
(\ref{thecon}) by a shift of $j+\half$ to $j$. Such shifts are
common in the identification of $R$ in terms of $j$ and $\theta$
for the fuzzy sphere. Heuristically we may say that each particle
occupies an area of $2\pi\theta$ in the previous fuzzy sphere
picture, while $2\pi\theta$ represents only an exclusion area
in the present picture.

\section{Equivalence to the Calogero-Sutherland particle model}
\label{Equivalence}

The dynamics of the above gauged matrix model with boundary field is
somewhat complicated to obtain, due to the nonlinear form of the action
in terms of $z$. It can be recast in more familiar form, however, upon
proper redefinition of variables. There are at least two different
ways to do this, leading, respectively, to a correspondence with
the well-known Calogero and Sutherland integrable systems.

To obtain the first correspondence, we define a new matrix $w$ as
\begin{equation}
w = 2R z (1+ z^\dagger z)^{-\frac{1}{2}}~.
\end{equation}
Since $z^\dagger z$ is a non-negative hermitian matrix, the square root
above is defined in the usual way in terms of the positive square roots
of the eigenvalues. In the scalar case the variable $w$ would represent
the ``chord length" of the particle from the north pole, along with its
azimuthal angle; the above is the proper matrix generalization. The inverse
transformation reads
\begin{equation}
z = w (4R^2 - w^\dagger w)^{-\frac{1}{2}}~.
\end{equation}

The lagrangian in terms of the new variable assumes the form
\begin{equation}
{\cal L} = i \frac{B}{2} \,\tr\left[ w^\dagger \dot{w}
- i A_0 ([w,w^\dagger ]- 2\theta) \right]
+ \Psi^\dagger (i \dot{\Psi} + A_o \Psi )~.
\end{equation}
This is identical to the lagrangian of the planar Chern-Simons matrix model.
The only remnant of the spherical topology is a constraint on $w$: from its
definition, it is clear that $w^\dagger w$ cannot have eigenvalues greater
that $2R$. That is, it must satisfy the constraint
\begin{equation}
w^\dagger w \le 4R^2
\label{constrw}
\end{equation}
($w w^\dagger$ has the same spectrum, apart, possibly, from zero modes
and will obey the same constraint.) The solution of this model, then, will
map to the solution of the corresponding truncation of the set of solutions of
the planar model.

Addition of single-particle potentials amounts to the addition
of a term $-\tr V(z,z^\dagger )$ in the action, which will map to a corresponding
planar potential ${\tilde V}(w,w^\dagger )$. Interestingly, the harmonic
oscillator potential on the plane 
\begin{equation}
{\tilde V} = \frac{1}{2} \omega^2 w^\dagger w
\label{Vharm}\end{equation}
maps (up to irrelevant constants) to a constant electric field in the vertical
direction on the sphere,
\begin{equation}
V = 2 R^2 \omega^2 (1+z^\dagger z)^{-1} \sim 
- R^2 \omega^2 \cos \theta
\end{equation}
corresponding to an electric field $E = \omega^2 R$.  The above
potential, as well as any potential that is a function of $w^\dagger
w$ alone, conserves the matrix $w^\dagger w$ over time and thus will
also preserve the constraint $w^\dagger w <4R^2$ on its eigenvalues,
once they are satisfied for the initial configuration. Arbitrary
potentials $\tilde V$ in $w$, however, can produce motions that exit
the constraint subspace. Such potentials always map to spherical
potentials $V$ with singular behavior at the south pole and can be
eliminated as unphysical.

As is well established, the planar model can be mapped to the Calogero
system of particles (for a review 
see \cite{Olshanetsky:1981dk, Polychronakos:1999sx}). 
The details of the mapping have been presented
before and will not be repeated here\footnote{For a direct mapping of
planar lowest Landau level anyons and quantum Hall states to the Calogero
system see \cite{Brink:1993sz, Azuma:1993ra}}. We just state the relevant
results.

The eigenvalues of the hermitian part of the matrix $w$, $X=
(w + w^\dagger)/2$, become particle coordinates, while the diagonal
elements of the antihermitian part $P = B (w-w^\dagger)/(2i)$, in the
basis where $X$ is diagonal, become their conjugate momenta. With the
harmonic potential (\ref{Vharm}) as the hamiltonian, the particle
hamiltonian is the standard Calogero hamiltonian with a one-body
harmonic oscillator potential and a two-body inverse-square potential
with strength $\theta^2$. The eigenvalues of the matrix $w^\dagger w$
are a set of classical ``pseudo-energies" $e_i$, whose sum is the
total energy. Higher conserved quantities of the model are written as
\begin{equation}
I_n = \tr (w^\dagger w)^n = \sum_{i=1}^N e_i^n ~~, n=1,2,\dots N~.
\label{consw}
\end{equation}

The constraint $[X,P]=iB\theta- i \Psi \Psi^\dagger$ implies that the
distance between the $e_i$ is at least $2\theta$. The ground state,
corresponds to the Calogero particles at rest in their equilibrium
position. It achieves the minimal values for the pseudo-energies, namely
$0, 2\theta, 4\theta, \dots , 2(N-1)\theta$. For the ground state to
exist, it must satisfy the constraint $w^\dagger w \le 4R^2$ so we
obtain
\begin{equation}
(N-1) \theta \le 2 R^2~.
\end{equation}
This constraint, and the ground state, are the ones obtained in the
previous section.  Excited states correspond to any other Calogero
particle configuration satisfying the constraints $e_i \le 4R^2$.

The second way to reduce the model amounts to a different
parametrization of the matrix $z$. Specifically, any matrix can be
parametrized in terms of a ``modulus" and ``phase" part, as
\begin{equation}
z= h U ~,
\end{equation}
where $h$ is a hermitian matrix and $U$ is a unitary matrix. They can
be expressed as
\begin{equation}
h = (z z^\dagger)^{\frac{1}{2}} ~~,~~~ U = (z z^\dagger)^{-\frac{1}{2}}\, z~.
\end{equation}
A further redefinition, more suitable for our purposes, is
\begin{equation}
H = (h^2 -1)(h^2 +1)^{-1} =  (z z^\dagger -1)(z z^\dagger +1)^{-1}~.
\end{equation}
In the scalar case the above variable would represent ($1/R$ times)
the vertical coordinate of the particle and is closely related
to the matrix $X_3$ defined before in (\ref{XXX}) (up to ordering).
The parametrization in
terms of $H$ and $U$, then, is the matrix version of the canonical
parametrization of the sphere in terms of $X_3$ and the azimuthal
angle $\phi$.

In terms of the new variables the lagrangian becomes (up to total derivatives)
\begin{equation}
{\cal L} = i BR^2 \,\tr\left[H {\dot U} U^{-1}
- i A_0 (H - U^{-1} H U - \theta /R^2) \right]
+ \Psi^\dagger (i \dot{\Psi} + A_o \Psi )~.
\label{UHlagrangian}
\end{equation}
This is identical to the lagrangian of the unitary matrix model
describing cylindrical quantum Hall states. Again, the only remnant of
the spherical topology is a constraint on $H$: from its definition, we
see that its eigenvalues have to be in the range $[ -1 , 1 ]$. The
solution of the model will, then, map to the solutions of the
cylindrical model truncated to the subspace satisfying the constraint.
Similar remarks on the inclusion of one-body potentials apply as in
the case of the previous mapping to the planar model.

The above matrix model is equivalent to the periodic modification of
the Calogero model, known as the Sutherland model. Again, we shall not
repeat the details but simply state the relevant facts.

The coordinates of the model can be taken as (the phase of) the
eigenvalues of $U$, representing particles on a circle of unit area,
while the diagonal elements of the matrix $P = BR^2 H$ in the basis where
$U$ is diagonal become particle momenta. The standard Sutherland
model obtains upon choosing $\frac{1}{2} \tr P^2$ as the
hamiltonian, and includes two-body inverse-sine-square interactions
with strength proportional to $\theta^2$.  The eigenvalues $p_i$
of $P$ become a set of ``pseudo-momenta" whose sum is the total
momentum $p$. The total momentum $p$, energy $E$ and a tower of higher
conserved quantities of the model are written as
\begin{equation}
I_n = \tr P^n = \sum_{i=1}^N p_i^n ~~~(I_1 = p, ~~ I_2 = 2E)~.
\end{equation}
The constraint $P - U^{-1} P U =B\theta- \Psi \Psi^\dagger$ implies
that the distance between the eigenvalues $p_i$ of $P$ is at least
$B\theta $. The ground state is when the particles are frozen to
their equilibrium position and corresponds to the $p_i$ assuming their
minimal values around zero, namely $-B\theta (N-1)/2 , \dots B\theta
(N-1)/2$. The eigenvalues of $H = P/BR^2$, then will range from $-\theta
(N-1)/2R^2$ to $\theta (N-1)/2R^2$. For this to be compatible with the
constraint $-1 \le H \le 1$ we must have
\begin{equation}
\theta \frac{N-1}{2R^2} \le 1~.
\end{equation}
which leads, again, to the same constraint as before.

In conclusion, the spherical matrix model can be described in terms of
the motion and degrees of freedom of either a Calogero or a Sutherland
particle system, with the initial data restricted so that they satisfy
the spherical constraint of the model.  All solutions can be obtained
in terms of the known classical solutions of these particle systems.
This will also be useful in the quantization of the model.

\section{Quantization}
\label{Quantization}

The full correspondence of the Chern-Simons matrix model to the
fractional quantum Hall system is born out at the quantum level, where
the quantization of the filling fraction and the charge of the
quasiholes naturally emerge.

The quantization of the present model will proceed along similar lines as the
quantization of the corresponding planar or cylindrical models. In fact, as was
demonstrated in the previous section, these models are equivalent up to the
special constraint that imposes the spherical topology.

To proceed, we choose the ``planar'' version of the model in terms of
the matrix $w$. Upon quantization, $w$, $w^\dagger$ and $\Psi$,
$\Psi^\dagger$ become matrices and vectors of oscillator ladder
operators respectively:
\begin{equation}
[ w_{a_a b_1} , w_{a_2 b_2}^\dagger ] = \frac{2}{B} \delta_{a_1 b_2}
\delta_{a_2 b_1} ~,~~~ [\Psi_a , \Psi_\beta^\dagger ] = \delta_{ab}~.
\end{equation} 
The states of the model are generated by the
repeated action of $w^\dagger$ and $\Psi^\dagger$ on the oscillator
ground state $|0>$ annihilated by all matrix elements of $w$ and
$\Psi$. Gauge invariance and the Gauss law constraint impose
restrictions on such states as well as the value of $\theta$. The
analysis is well known and we again state the results.

The Gauss law constraint quantum mechanically becomes a statement on 
the allowed
representations of the generators of the conjugation symmetry of $w$, 
$w^\dagger$.
Group theory constraints require the quantization condition
\cite{Polychronakos:1991bx, Nair:2001rt, Polychronakos:2001mi}
\begin{equation} 
\theta B = k ~, ~~~k=0,1,2,\dots~.
\end{equation}
This will be related to the quantization of the filling fraction.

The implementation of the spherical constraint (\ref{constrw}) is more
subtle and will lead to another quantization condition. Quantum
mechanically it is unclear to what the eigenvalues of this matrix
would correspond. One way to resolve the issue is to work, instead,
with the invariants of the model $I_n$ as given in (\ref{consw}). The
restriction $e_i \le 2R$ imposes corresponding restrictions on $I_n$.

Quantum mechanically, $I_n$ are the corresponding commuting quantum 
integrals of the
model. As is known from the matrix model, or its corresponding 
Calogero system, such
integrals can be written in terms of a set of quantum mechanical 
``pseudo-excitation
numbers" $m_i$, $i=1,2,\dots N$
\begin{equation}
I_n = \sum_{i=1}^N \left(\frac{2m_i}{B}\right)^n~.
\end{equation}
The $m_i$ are non-negative integers satisfying the constraint
\begin{equation}
m_{i+1} - m_i \le k+1 ~,~~~i=1,2,\dots N-1~.
\label{mcon}
\end{equation}
This is very similar to the corresponding classical constraint on the
eigenvalues of $w^\dagger w$ upon putting $m_i = B e_i /2$ and $k =
\theta B$, the only difference being the shift of $k$ to $k+1$. This
is the well-known level shift of the coefficient of the Chern-Simons
term, leading to the ``fermionization'' of the matrix model and the
renormalization of the filling fraction.

The spherical constraint, now, as expressed in terms of $I_n$ can be
mapped back to constraints on the $m_i$. Since $m_i$ enters exactly as
$B e_i /2$ in the expressions for $I_n$, this constraint
simply reads
\begin{equation}
m_i \le 2BR^2~.
\label{Bcon}
\end{equation}

One can picture the $m_i$ as $N$ points on a non-negative linear
integer lattice, with the extra constraint (\ref{mcon}) that they have
to be at least $\ell = k+1$ lattice units apart.  The constraint
(\ref{Bcon}) means that they must also be less than $2 B R^2$. In
fact, closer scrutiny reveals that $2BR^2$ must fall {\it exactly} on
a lattice point. The reason is similar to the one that makes the
spectrum of $m_i$ start from zero: the point $w=0$ corresponds to the
north pole of the sphere and $w$ must have a zero mode there to
prevent the appearance of unphysical states lying ``north'' of the pole.
Similarly, the point at $w^\dagger w = 2R$, where the spherical
constraint is saturated, corresponds to the south pole of the sphere
and there must be a state saturating it to prevent the creation of
unphysical states ``south'' of the south pole.  We conclude that there
must be a lattice point $m$ satisfying
\begin{equation}
2BR^2 = m~.
\end{equation}
Written in the form
\begin{equation}
B 4\pi R^2 = 2\pi m
\end{equation}
the above recovers the standard monopole quantization for the sphere,
restricting the total flux to an integer number of flux quanta. The
integer $m$ is the monopole number.  (The above discussion may strike
the reader as somewhat heuristic. In fact, it parallels the discussion
of the corresponding scalar system of a massless particle on the
sphere where the monopole quantization is also recovered. A more
careful treatment would involve adding a proper quadratic kinetic
term for the
model as a regulator, but we will forgo any such elaboration here.)

In conclusion, we see that the quantum model contains two quantized
parameters, the monopole number $m$ and the ``level'' number $\ell =
k+1$. The ground state of the model corresponds to the
quasi-occupation numbers having their lowest values, namely $0, \ell,
2\ell , \dots (N-1)\ell$. For this to exist we must have
\begin{equation}
(N-1)\ell \le m~.
\end{equation}
The saturation value $(N-1)\ell =m$ corresponds to a fully filled
sphere. Given that the lowest Landau level of a monopole $m$ magnetic
field on the sphere level is in the $j=\frac{m}{2}$ angular momentum
representation, it contains $L = 2j+1 = m+1$ states (corresponding
to the lattice points $0,1,2,\dots m$).  The ratio $N/L$
for the saturated state above, then, is
\begin{equation}
\nu = \frac{L + \ell -1}{\ell L}~.
\end{equation}
In the limit $L,N \to \infty$ we recover a filling fraction
\begin{equation}
\nu = \frac{1}{\ell} = \frac{1}{k+1}~.
\end{equation}
This confirms that the inverse filling fraction is given by the
renormalized coupling constant $\ell$, rather than $k$.

In the non-saturated case the above ground state forms a ``droplet'' of
$\nu = 1/\ell$ quantum Hall fluid filling a northern section of the
sphere. The degeneracy $G$ of the possible Hall states is given by the
number of possible distinct ways to place the $N$ integers $m_i$ on
the lattice $0,1,\dots m$ respecting the constraints $m_{i+1} -m_i \ge
\ell$. It can be shown that this degeneracy is
\begin{equation}
G = \frac{[m-(N-1)\ell +N]!}{[m-(N-1)\ell]! N!}~.
\end{equation}
This exactly matches the degeneracy of the corresponding $\nu =
1/\ell$ Haldane states on a sphere with a monopole of strength $m$
at the center.\footnote{
For comparison, our monopole number $m$ is Haldane's $2S$ and our
level $\ell$ is Haldane's $m$.}. We have
therefore established the correspondence of this model with the
fractional quantum Hall states on the sphere.

The above analysis can be repeated in terms of the unitary model
$H,U$. The treatment is along the same lines, leading to a similar
picture for the states and the same quantization conditions for
the monopole and the filling fraction.

\section{Concluding remarks}
\label{remarks}
We have presented a model that realizes the fractional quantum Hall
system on a sphere as a matrix model, extending previous work on the
plane and cylinder, and demonstrated that it reproduces the Hilbert
space of particles as predicted by the standard Haldane states.

Apart from representing a technical advance in the quantum Hall matrix
model technology, this also puts the issue of mapping of states and
filling fraction on a firmer footing. The filling fraction in the
present model is the ratio of two integers, $N$ and $L=m+1$, and
therefore is unambiguously defined and shown to equal $1/(k+1)$,
confirming the renormalization of the level number to $k+1$. In
the planar case, in the absence of an exact operator mapping for
the density, this was somewhat open to interpretation.

An interesting open issue is the derivation of the exact
angular momentum operators
$J_\pm , J_3$ in the quantum case. These would be given by some
quantum ordering of the classical expressions (\ref{Jclass})
and would satisfy the full quantum $SU(2)$ algebra. The identification
of these operators would essentially resolve the question of
the mapping of Hilbert space states
in the Hall system and the matrix model. Indeed, the ground state
is unambiguously defined as the fully filled circular droplet,
just as in the planar case, but excited states are degenerate
(in terms of their energy or expectation value of $x_3$ coordinates)
and the exact
mapping is ambiguous. These states, however, group into distinct
irreducible representations of the rotation algebra of the sphere
and are uniquely identified by their corresponding quantum
numbers. Identifying this algebra would, then, afford us the
full Hilbert space map.

Several other issues remain for further investigation. The model is
essentially a (gauged) matrix generalization of the corresponding
scalar model of a particle on the sphere.  It is known that the scalar
model derives from a Kirillov action over an SU(2) group manifold,
upon proper reduction to a quotient space SU(2)/U(1). There should be
a corresponding Kirillov-type action for the full matrix model with
interesting mathematical properties, whose derivation remains an open
issue.  Further, Calogero-type models have been mapped to
two-dimensional gauge theory and three-dimensional topological gauge
theories \cite{Gorsky:1993pe, Minahan:1993mv}.  The above spherical
model, then, should correspond to a topological field theory, as it
has a finite-dimensional Hilbert space, whose derivation is another
interesting problem. Finally, the physical question of properly
incorporating spin and composite filling fractions in the matrix model
is still an open issue.

\section*{Acknowledgments}
This work was supported in part by the~ U.S.~ Department~ of~ Energy~
under ~Contract Number DE-FG02-91ER40651-TASK B and by the National
Science Foundation under grant PHY-0353301.

\appendix
\section*{Appendix}
\label{appendix}
In this appendix we prove that the charges $J_\pm , J_3$
satisfy the $su(2)$ Lie algebra\footnote{More precisely, we have computed
  the algebra of Poisson brackets. Computing the quantum commutator of 
  these charges is an interesting open problem.}
\begin{equation}
\left\{
\begin{array}{lll}
i \{J_3, J_{\pm} \} & = & \pm  J_{\pm}~, \\
i \{J_+, J_- \} & = & 2 J_3~.
\end{array} 
\right. \label{poissonLie}
\end{equation}

We shall use the parametrization of $z$ introduced in section
(\ref{Equivalence})
\begin{equation}
  U =(zz^\dagger)^{-\half} z~, ~~~H=(z z^\dagger-1)(z z^\dagger +1)^{-1}~.
\end{equation}
Then the charges are given by
\begin{equation}
\left\{
\begin{array}{lll}
J_3  & = & BR^2 \, \tr(H)~, \\
J_{+} & = &  BR^2 \, \tr(\sqrt{1-H^2}\,U)~, \\
J_{-} & = &  BR^2 \, \tr(\sqrt{1-H^2}\,U^{-1})~.
\end{array}
\right.
\end{equation}
From the kinetic term of the
lagrangian~\rref{UHlagrangian} given by ${\cal L} = i BR^2 \tr(H\,\dot{U}
U^{-1})$ we read off the Poisson brackets
\begin{equation}
\left\{
\begin{array}{lll}
 \{U_1^{-1}H_1, U_2\} & = & \frac{1}{i BR^2}  T_{12}~, \\
\{U_1, U_2\} & = & 0 ~,\\
 \{U_1^{-1}H_1,U_2^{-1}H_2\} & = & 0~.
\end{array}
\right.\label{poisson}
\end{equation}
Here, as in the main text, the subscripts denote the space in which
the matrices operate and $(T_{12})^{i_1 i_2}_{j_1 j_2} =
\delta_{j_1}^{i_2} \delta_{j_2}^{i_1}$ is an operator exchanging the
two spaces.
These relations~\rref{poisson} can be written equivalently as
\begin{equation}
\left\{
\begin{array}{lll}
\{H_1, H_2\} & = &  \frac{1}{i BR^2}(T_{12} H_2 - T_{12} H_1)~, \\
\{U_1, U_2\} & = & 0~, \\
\{H_1, U_2\} & = &  \frac{1}{i BR^2} T_{12} U_2~.
\end{array}
\right.
\end{equation}
Using the identity
\begin{equation}
\{H_1, f(H_2)\} =  \frac{1}{i BR^2}
[T_{12} f(H_2) -
T_{12} f(H_1)]~,\label{identity}
\end{equation}
where $f$ is an arbitrary function, we can check that
\begin{equation}
\{H_1, f(H_2)U_2\} =  \frac{1}{i BR^2}
T_{12} f(H_2) U_2
\end{equation}
and after tracing on both subspaces we get
\begin{equation}
 i \{BR^2\tr(H), BR^2\tr(f(H)U)\} =  BR^2\tr(f(H)U)~.\label{pppm1}
\end{equation}
In the same way we can show that
\begin{equation}
 i\{BR^2\tr(H), BR^2\tr(f(H)U^{-1})\} = - BR^2 \tr(f(H)U^{-1})~.\label{pppm2}
\end{equation}
Then if we set $f(H)=  \sqrt{1-H^2}$ in \rref{pppm1} and \rref{pppm2} we
obtain the first line in \rref{poissonLie}\,.

Deriving the last line in \rref{poissonLie} is a little more difficult and the
specific form of the function $f$ is needed. First let us assume that
$f$ has the expansion
\begin{equation}
f(H)=\sum_{k=0}^{\infty} a_k H^k ~.
  \label{expansion}
\end{equation}
Then for an arbitrary matrix valued function $A$ we have
\begin{equation}
  \{A_1, f(H_2)\}
=
\sum_{m,n=0}^{\infty} a_{m+n-1} \, H_2^m \{A_1,H_2\} H_2^n~,
\label{expand}
\end{equation}
where $a_{-1} = 0$ is assumed to vanish.

As above we start with $\{f(H_1)U_1, f(H_2)U_2^{-1}\}$ and after
expanding using \rref{identity} and \rref{expand} and tracing we obtain
\begin{equation}
\{
 \tr(f(H)U), \tr(f(H)U^{-1})\}
  = \!
\sum_{m,n=0}^{\infty} i  a_{m+n-1}
\tr(f(H)H^mU^{-1}H^n U +
f(H)H^m U H^n U^{-1})~. \nonumber
\end{equation}
After further expanding $f$\,, changing the order of
summation and some index redefinition, we obtain
\begin{equation}
  \{ \tr(f(H)U), \tr(f(H)U^{-1})\}
= \frac{i}{ B R^2}\sum_{m,n=0}^{\infty}
\alpha_{m+n+1}
\tr(H^mU^{-1}H^nU)~,
\end{equation}
where $\alpha_{t} = \sum_{k=0}^{t} a_{t-k} a_{k}$\,. The convolution
coefficients can be recognized as the expansion coefficients of
$f^2(H)=1-H^2$ so only $\alpha_0= 1$ and $\alpha_2= -1$ are
nonvanishing and furthermore only $\alpha_2$ contributes to the sum. 
Finally we obtain
\begin{equation}
 i  \{ B R^2 \tr(\sqrt{1-H^2})U), B R^2\tr(\sqrt{1-H^2}U^{-1})\}
= 2 BR^2\tr(H)~.
\end{equation}
This completes the proof of the relations \rref{poissonLie}\,. In the
derivation it was essential to be able to expand the function $f(H)$
as in \rref{expansion}.
This is always possible in our case since all
the  eigenvalues of $H$ have  absolute values less than one. The
$SO(3)$ symmetry is not a symmetry of the general unitary model
and the restrictions on the integrals of motions are essential.


\begin{thebibliography}{99}



\bibitem{Susskind:2001fb}
  L.~Susskind,
  ``The quantum Hall fluid and non-commutative Chern Simons theory,''
  arXiv:hep-th/0101029.


\bibitem{Seiberg:1999vs}
  N.~Seiberg and E.~Witten,
   ``String theory and noncommutative geometry,''
  JHEP {\bf 9909}, 032 (1999)
  [arXiv:hep-th/9908142].


\bibitem{Harvey:2001yn}
  J.~A.~Harvey,
  ``Komaba lectures on noncommutative solitons and D-branes,''
  arXiv:hep-th/0102076.


\bibitem{Douglas:2001ba}
  M.~R.~Douglas and N.~A.~Nekrasov,
  ``Noncommutative field theory,''
  Rev.\ Mod.\ Phys.\  {\bf 73}, 977 (2001)
  [arXiv:hep-th/0106048].


\bibitem{Szabo:2001kg}
  R.~J.~Szabo,
  ``Quantum field theory on noncommutative spaces,''
  Phys.\ Rept.\  {\bf 378}, 207 (2003)
  [arXiv:hep-th/0109162].


\bibitem{Polychronakos:2001mi}
  A.~P.~Polychronakos,
  ``Quantum Hall states as matrix Chern-Simons theory,''
  JHEP {\bf 0104}, 011 (2001)
  [arXiv:hep-th/0103013].


\bibitem{Hellerman:2001rj}
  S.~Hellerman and M.~Van Raamsdonk,
  ``Quantum Hall physics equals noncommutative field theory,''
  JHEP {\bf 0110}, 039 (2001)
  [arXiv:hep-th/0103179].


\bibitem{Karabali:2001xq}
  D.~Karabali and B.~Sakita,
  ``Chern-Simons matrix model: Coherent states and relation to Laughlin
  wavefunctions,''
  Phys.\ Rev.\ B {\bf 64}, 245316 (2001)
  [arXiv:hep-th/0106016];
  ``Orthogonal basis for the energy eigenfunctions of the Chern-Simons  matrix
  model,''
  Phys.\ Rev.\ B {\bf 65}, 075304 (2002)
  [arXiv:hep-th/0107168].

\bibitem{Hansson:2001kt}
T.~H.~Hansson, J.~Kailasvuori and A.~Karlhede, 
``Charge and current in the quantum Hall matrix model,'' 
Phys.\ Rev.\ B {\bf  68}, 035327 (2003).


\bibitem{Cappelli:2004xk}
A.~Cappelli and M.~Riccardi, 
``Matrix model description of Laughlin Hall states,'' 
arXiv:hep-th/0410151.

\bibitem{Polychronakos:2001uw}
  A.~P.~Polychronakos,
  ``Quantum Hall states on the cylinder as unitary matrix Chern-Simons
  theory,''
  JHEP {\bf 0106}, 070 (2001)
  [arXiv:hep-th/0106011].


\bibitem{Morariu:2001qa}
  B.~Morariu and A.~P.~Polychronakos,
``Finite noncommutative Chern-Simons with a Wilson 
     line and the quantum Hall effect,''
  JHEP {\bf 0107}, 006 (2001)
  [arXiv:hep-th/0106072].

\bibitem{Chen:2003ic}
  Y.~X.~Chen, M.~D.~Gould and Y.~Z.~Zhang,
  ``Finite matrix model of quantum Hall fluids on S**2,''
  arXiv:hep-th/0308040.


\bibitem{Berenstein:2004hw}
  D.~Berenstein,
  ``A matrix model for a quantum Hall droplet with manifest particle-hole
  symmetry,''
  Phys.\ Rev.\ D {\bf 71}, 085001 (2005)
  [arXiv:hep-th/0409115].

\bibitem{Ghodsi:2005ks}
  A.~Ghodsi, A.~E.~Mosaffa, O.~Saremi and M.~M.~Sheikh-Jabbari,
  ``LLL vs. LLM: Half BPS sector of N = 4 SYM equals to quantum Hall system,''
  arXiv:hep-th/0505129.

\bibitem{Laughlin:1983fy}
  R.~B.~Laughlin,
  ``Anomalous Quantum Hall Effect: An Incompressible Quantum Fluid With
  Fractionally Charged Excitations,''
  Phys.\ Rev.\ Lett.\  {\bf 50} (1983) 1395.
 

\bibitem{Haldane:1983xm}
  F.~D.~M.~Haldane,
  ``Fractional Quantization Of The Hall Effect: A Hierarchy Of Incompressible
  Quantum Fluid States,''
  Phys.\ Rev.\ Lett.\  {\bf 51} (1983) 605.



\bibitem{Madore:1991bw}
J.~Madore,
``The Fuzzy sphere,''
Class.\ Quant.\ Grav.\ {\bf 9}, 69 (1992)


\bibitem{Carow-Watamura:1996wg}
U.~Carow-Watamura and S.~Watamura,
``Chirality and Dirac operator on noncommutative sphere,''
Commun.\ Math.\ Phys.\ {\bf 183}, 365 (1997)
[hep-th/9605003];
``Noncommutative geometry and gauge theory on fuzzy sphere,''
Commun.\ Math.\ Phys.\ {\bf 212}, 395 (2000)
[hep-th/9801195];
``Differential calculus on fuzzy sphere and scalar field,''
Int.\ J.\ Mod.\ Phys.\ A {\bf 13}, 3235 (1998);
[q-alg/9710034]


\bibitem{Klimcik:1997mg}
C.~Klimcik,
``Gauge theories on the noncommutative sphere,''
Commun.\ Math.\ Phys.\ {\bf 199}, 257 (1998)
[hep-th/9710153];


\bibitem{Gratus:1997rq}
J.~Gratus,
``An introduction to the noncommutative sphere and some extensions,''
q-alg/9710014;

\bibitem{Morariu:1997vu}
  B.~Morariu,
  ``Path integral quantization of the symplectic leaves of the SU(2)
  Poisson-Lie group,''
  Int.\ J.\ Mod.\ Phys.\ A {\bf 14}, 919 (1999)
  [arXiv:physics/9710010].


\bibitem{Nair:2000ii}
  V.~P.~Nair and A.~P.~Polychronakos,
  ``Quantum mechanics on the noncommutative plane and sphere,''
  Phys.\ Lett.\ B {\bf 505}, 267 (2001)
  [arXiv:hep-th/0011172].


\bibitem{Morariu:2004aw}
  B.~Morariu,
  ``Strings, dipoles and fuzzy spheres,''
  arXiv:hep-th/0408018.




\bibitem{Calogero:1970nt}
  F.~Calogero,
  ``Solution Of The One-Dimensional N Body Problems With Quadratic And/Or
  Inversely Quadratic Pair Potentials,''
  J.\ Math.\ Phys.\  {\bf 12} (1971) 419.


\bibitem{Sutherland:1971kq}
  B.~Sutherland,
  ``Exact Results For A Quantum Many Body Problem In One-Dimension,''
  Phys.\ Rev.\ A {\bf 4} (1971) 2019;
  ``Exact Results For A Quantum Many Body Problem In One-Dimension. 2,''
  Phys.\ Rev.\ A {\bf 5} (1972) 1372.



\bibitem{Bahcall:1991an}
  S.~Bahcall and L.~Susskind,
  ``Fluid dynamics, Chern-Simons theory and the quantum Hall effect,''
  Int.\ J.\ Mod.\ Phys.\ B {\bf 5}, 2735 (1991).



\bibitem{Dunne:1989hv}
  G.~V.~Dunne, R.~Jackiw and C.~A.~Trugenberger,
  ``'Topological' (Chern-Simons) Quantum Mechanics,''
  Phys.\ Rev.\ D {\bf 41}, 661 (1990).



\bibitem{Binetruy:1984yx}
  P.~Binetruy and M.~K.~Gaillard,
  ``Temperature Corrections In The Case Of Derivative Interactions,''
  Phys.\ Rev.\ D {\bf 32}, 931 (1985).


\bibitem{Gaillard:1997zr}
  M.~K.~Gaillard and B.~Zumino,
  ``Self-duality in nonlinear electromagnetism,''
  arXiv:hep-th/9705226;
  ``Nonlinear electromagnetic self-duality and Legendre transformations,''
  arXiv:hep-th/9712103.


\bibitem{Alexanian:2000uz}
  G.~Alexanian, A.~Pinzul and A.~Stern,
  ``Generalized Coherent State Approach to Star Products and Applications to
  the Fuzzy Sphere,''
  Nucl.\ Phys.\ B {\bf 600}, 531 (2001)
  [arXiv:hep-th/0010187].


\bibitem{Iso:2001mg}
  S.~Iso, Y.~Kimura, K.~Tanaka and K.~Wakatsuki,
  ``Noncommutative gauge theory on fuzzy sphere from matrix model,''
  Nucl.\ Phys.\ B {\bf 604}, 121 (2001)
  [arXiv:hep-th/0101102].


\bibitem{Olshanetsky:1981dk}
  M.~A.~Olshanetsky and A.~M.~Perelomov,
  ``Classical Integrable Finite Dimensional Systems Related To Lie Algebras,''
  Phys.\ Rept.\  {\bf 71} (1981) 313.


\bibitem{Polychronakos:1999sx}
  A.~P.~Polychronakos,
  ``Generalized statistics in one dimension,''
Les Houches Lectures, Summer 1998,
  arXiv:hep-th/9902157.

\bibitem{Brink:1993sz}
L.~Brink, T.~H.~Hansson, S.~Konstein and M.~A.~Vasiliev, 
``The Calogero model: Anyonic representation, fermionic extension 
and supersymmetry,'' 
Nucl.\ Phys.\ B {\bf 401}, 591 (1993) [arXiv:hep-th/9302023]. 


\bibitem{Azuma:1993ra}
  H.~Azuma and S.~Iso,
  ``Explicit relation of quantum hall effect and Calogero-Sutherland model,''
  Phys.\ Lett.\ B {\bf 331}, 107 (1994)
  [arXiv:hep-th/9312001].

\bibitem{Polychronakos:1991bx}
A.~P.~Polychronakos, 
``Integrable systems from gauged matrix models,'' 
Phys.\ Lett.\ B {\bf 266}, 29 (1991). 


\bibitem{Nair:2001rt}
  V.~P.~Nair and A.~P.~Polychronakos,
  ``On level quantization for the noncommutative Chern-Simons theory,''
  Phys.\ Rev.\ Lett.\  {\bf 87}, 030403 (2001)
  [arXiv:hep-th/0102181].



\bibitem{Gorsky:1993pe}
  A.~Gorsky and N.~Nekrasov,
  ``Hamiltonian systems of Calogero type and two-dimensional Yang-Mills
  Nucl.\ Phys.\ B {\bf 414}, 213 (1994)
  [arXiv:hep-th/9304047];
  ``Quantum integrable systems of particles as gauge theories,''
  Theor.\ Math.\ Phys.\  {\bf 100}, 874 (1994)
  [Teor.\ Mat.\ Fiz.\  {\bf 100}, 97 (1994)].

\bibitem{Minahan:1993mv} 
J.~A.~Minahan and A.~P.~Polychronakos, 
``Interacting fermion systems from two-dimensional QCD,''
 Phys.\ Lett.\ B {\bf 326}, 288 (1994) [arXiv:hep-th/9309044].

\end{thebibliography}
\end{document}